\documentclass[twocolumn]{aa}
\usepackage{txfonts}
\usepackage{graphicx}

\def\etal{{\it et~al. }} 

\begin{document}

\title{Are retrograde resonances possible in multi-planet systems?}
\author{Julie Gayon \& Eric Bois}
\institute{Nice Sophia-Antipolis University, CNRS, Observatoire de la C\^ote
d'Azur, Laboratoire Cassiop\'ee, B.P. 4229, F-06304 Nice Cedex 4, France\\
\email{Julie.Gayon@oca.eu.fr - Eric.Bois@oca.eu}}

\authorrunning{Gayon \& Bois}
\titlerunning{Are retrograde resonances possible in multi-planet systems?}

\date{}

\abstract
{Most multi-planetary systems are characterized by hot-Jupiters close to their
central star, moving on eccentric orbits. From a dynamical point of view,
compact multi-planetary systems form a specific class of the general {\it
N}-body problem (where $N\ge3$). Moreover, extrasolar planets are found in
prograde orbits about their host star, and often in mean
motion resonances (MMR).} 
{In a first step, we study theoretically a new stabilizing mechanism suitable
for compact two-planet systems. This mechanism involves
counter-revolving orbits forming a retrograde MMR. In a second step, we
investigate the feasibility of planetary systems hosting counter-revolving
planets. Dynamical stability, observations, and formation processes of these
systems are analyzed and discussed.}
{To characterize the dynamical behavior of multi-dimensional planetary 
systems, we apply our
technique of global dynamics analysis based on the MEGNO
indicator (Mean Exponential Growth factor of Nearby Orbits) that provides the
fine structure of the phase space. In a few cases of possible counter-revolving
configurations, we carry out new fits to the observations using the Pikaia 
genetic algorithm. A statistical study of the stability in the neighborhood of
different observed, planetary-systems is completed using a Monte-Carlo
method.}  
{We analyse the observational data for the HD\thinspace73526 planetary system
and find that counter-revolving configurations may be consistent with the
observational data.
We highlight the fine and characteristic structure of
retrograde MMRs. We demonstrate that retrograde resonances open a family of
stabilizing mechanisms involving new apsidal precession behaviors.} 
{Considering two possible formation mechanisms (free-floating planet and 
Slingshot model), we conclude that counter-revolving configurations are 
feasible.}


\maketitle

\section{Introduction}
At present, 271 extrasolar planets have been detected around 233 stars (both
solar and non-solar type)\footnote{January, the $\textrm{11}^\textrm{th}$, 
2008. http://exoplanet.eu/catalog.php}. Among them, there are 25 
multiple-planet systems~: 17 two-planet systems (e.g. HD$\thinspace$82943, 47
UMa, HD$\thinspace$108874, HD$\thinspace$128311), 6 three-planet systems
(e.g. $\upsilon$ And, HD$\thinspace$69830, Gliese 876,
Gliese 581), 1 four-planet system (HD$\thinspace$160691) and more recently 1
five-planet system ($55$ CnC). 
Observations indicate that Mean Motion Resonances (MMR) frequently occur for 
planets of multiple-planet systems~: Gliese 876 (e.g. Rivera \etal 2005), 
HD$\thinspace$82943 (e.g. Ji \etal 2003, Mayor \etal 2004) and
HD$\thinspace$128311 (Vogt \etal 2005) are in 2:1 MMR, HD$\thinspace$202206 is
in 5:1 MMR (Correia \etal 2005), while 47 UMa is close to a 7:3 (Fischer \etal
2002) or 8:3 commensurability (Fischer \etal 2003). 

This work is devoted to compact multi-planetary systems, characterized by (a) 
giant Jupiter-like planets found close to their central star, and (b) high
eccentricities. These two peculiarities lead to strong gravitational
interactions between the planets and may result in an unstable, dynamical
behavior. However, we\linebreak observe many such planetary systems suggesting that they 
are stable, and raising the question of why they are stable. From a dynamical 
point of view, compact multi-planetary systems form a specific class of the
general N-body problem (with $N \ge 3$)  whose analytical solutions are not
necessarily known. A stability analysis of planetary systems, using numerical
methods to explore multi-dimensional parameter space, typically leads to
stability maps in
which rare islands of stability can be identified amidst large chaotic zones.  
The underlying mechanisms for these stability zones must be identified.

In 2002, Kiseleva-Eggleton \etal (2002) showed that the currently-published, 
orbital parameters place the planetary systems HD 12661, HD 38529, 
HD 37124, and HD$\thinspace$160691 in very different situations from the point
of view of dynamical distribution. Since this first study of the comparative
stability of multi-planetary systems, many studies\linebreak have been carried out
in this direction. The role of the orbital mean motion resonances,
in particular with a 2:1 ratio, has been intensively studied by
several research groups (for example Hadjidemetriou 2002; Lee \& Peale 2002,
2003; Bois \etal 2003; Ji \etal 2003; Ferraz-Mello \etal 2005b; Psychoyos \& 
Hadjidemetriou 2005; Beaug\'e \etal 2006). 
As a result, it has been discovered that an extrasolar planetary system, even
with large planetary masses  
and eccentricities, can be stable if planetary orbits are close to stable,
resonant, and periodic orbits. It has also been established (see e.g. Chiang \&
Murray 2002; Lee \& Peale 2002; Libert \& Henrard 2006) that orbits in a large
number of compact multi-planet systems, are locked in Apsidal Synchronous
Precessions (ASP hereafter), i.e. that the apsidal lines precess, on average, 
at the same rate\footnote{For the study of 3-D, full 
3-body problems, we introduced the terminology ASP, for
expressing that the apsidal lines precess on average in a 3-D space at the 
same rate: see Bois 2005, Bois \etal 2005. We note that in the planar
case this phenomenon is also called ``apsidal corotation''
(ACR; Beaug\'e 
\etal 2003). In a number of papers, one may also find the
  incorrect expression ``apsidal secular resonances'' (ASR). 
ACR and ASP are in general not true secular resonances, as highlighted by
Ferraz-Mello \etal (2005a).}. 
A solution involving\linebreak both MMRs and ASP describes well the stability of
eccentric, compact multi-planetary systems, but may not however be unique. We
note, for example, that other multi-planetary systems have been
found to be mainly
controlled by secular dynamics (cf. Michtchenko \etal 2006; Libert
\& Henrard 2006; Ji \etal 2007). In the present paper, we illustrate
theoretically that other mechanisms can in addition provide the stability
in multi-planetary systems.

In the case of the HD$\thinspace$73526 system (2:1 MMR), Tinney \etal (2006) 
found stability over 1 Myr. Based on 
the analytical\linebreak classification of Hadjidemetriou (2002) established
according to
a hierarchy of masses and eccentricities, this system could
instead be classified as
unstable. Hadjidemetriou's classification may however be too
general to disprove the stability found by Tinney \etal (2006). 
Be that as it may,
we use the same data as Tinney \etal (see Table \ref{tab_param}) and our numerical
method is outlined in the following section. 
Exploring the stability of the HD$\thinspace$73526 system in
orbital parameter space, we find large chaotic regions.
We find that the published data can even be described by a chaotic behavior. We note
however that Tinney \etal (2006) used a different definition of stability.\footnote{In 
the paper of Tinney \etal (2006), the claim of stability is 
obtained from the dynamical behaviors of the resonant angles related to the 
2:1 MMR, rather than by characterizations of 
quasi-periodicity of the orbital solution. Besides, the notion of stability is 
only presumed to be acquired by the simple absence of planet
ejection. We use instead the usual definition of stability related to
quasi-periodicity (see Section \ref{method}) and suitable for conservative
dynamical systems.}
Of course, we cannot exclude that the observational data were insufficient
to allow a reliable orbital fit or the fit itself was not\linebreak adequate.
On the other hand, it is also possible that the underlying assumption
of two prograde orbits is wrong. When placing\linebreak one of the two planets on a 
retrograde orbit (which forms a system with 
counter-revolving planets), the stability region\linebreak becomes very large. We will 
show below
that this does not imply\linebreak that the orbital fit is consistent with this 
stability zone. It implies\linebreak that, in the neighborhood of the
observational point, we can\linebreak theoretically find stable solutions 
for counter-revolving configurations. To distinguish between two 
resonance cases when both planets are in prograde orbits, or when one planet
is on a retro\-grade orbit, we call them prograde and retrograde resonances, 
respectively. 

Presently, all known extrasolar planets in multiple systems
are believed to revolve in the same direction about 
their corresponding central star. Most fitted, orbital elements
are derived by assuming prograde orbits. This is expected according to current\linebreak 
theories for planetary formation in a circumstellar disk.
In order to obtain a planet in retrograde resonance, an additional event
is necessary such as violent, dynamical evolution of the planetary system,
or a capture of the retrograde planet. In our Solar System,
comets and the planetary satellites of Neptune, Saturn and Jupiter are known
to have retrograde orbits. It is, therefore,\linebreak important to investigate the
stability of exoplanetary systems with a 
retrograde planet in particular if the observations do not yield a stable 
system when assuming all planets on prograde\linebreak orbits.

In Section \ref{method}, we present our method of global
dynamics analysis. We show that there exists theoretically 
initial conditions in the vicinity of observational data such
that stability is only possible for a counter-revolving configuration.
It raises the question of whether such a configuration is consistent with
the observational\linebreak data of a given system (Section \ref{obs}). In
Section \ref{stat}, we focus on the statistical occurence of stable
solutions related to both prograde and retrograde resonances.
This statistical approach is applied to three
systems in 2:1 MMR and two systems in 5:1 MMR. If such systems harboring
counter-revolving planets exist, we must also consider how they form: we 
discuss
this issue in Section \ref{form}. By analyzing the parameter space in the
vicinity of the best-fit of the HD\thinspace73526 planetary system, we
highlight the fine structure of the 2:1 retrograde resonance
(Sections \ref{structure} and \ref{evidence}), and the nature of associated 
apsidal precessions (Section \ref{apsidal}).  
In addition, we complete an analoguous study for a
theoretical system in 5:1 retrograde MMR (Section \ref{HD16}).

\section{Method}\label{method}
In order to explore the stability in the parameter space of known 
exoplanetary system in the case of retrograde resonance,
we use the MEGNO (Mean Exponential Growth
factor of Nearby Orbits) method proposed by Cincotta \& Sim\`o (2000). This 
method provides relevant information on the global dynamics of
multi-dimensional, Hamiltonian systems and the fine structure\linebreak of their phase 
space (Cincotta \etal 2002). It simultaneously yields a good estimate of
the Lyapunov Characteristic Numbers (LCN) with a comparatively
small computational\linebreak effort (Cincotta \& Giordano 2000). 
It provides a clear picture
of resonance structures, location of stable and unstable periodic\linebreak orbits, as
well as a measure of hyperbolicity in chaotic domains\linebreak (i.e. the rate of
divergence of unstable orbits). Using the MEGNO
technique, we have built the MIPS (Megno Indicator for
Planetary Systems) package specifically devoted to studying multi-dimensional
planetary systems and their conditions of\linebreak dynamical stability. We use the 
property of stability in the Poisson sense:  stability
is related to the preservation of a neighborhood related to the
initial position of the trajectory. Moreover, in the
Poincar\'e-Lyapunov sense applied to conservative systems, when 
quasi-periodic orbits remain confined within certain limits, they are
called {\it stable}. We note that {\it chaotic}, in the Poincar\'e 
sense means that the dynamical behavior is not quasi-periodic (according to
the conventional definition used for conservative dynamical systems)
and does not necessarily mean that the system will disintegrate during 
limited period of time. 
We have already successfully applied the MEGNO technique with
the MIPS package\linebreak to the study of dynamical stability of extrasolar
planetary\linebreak 
systems in a series of previous papers (see e.g. Bois \etal 2003, 2004). 
In the MIPS package, let us note that the $i_{b,c}$ inclination parameters refer
to the dynamical, orbital-element independent of the $sin\,i_l$ line-of-sight
inclination factor.\footnote{In the present paper, masses of
planets remain untouched whatever the mutual inclinations may be. 
Our reference frame is related to the planetary system itself,
then dynamically autonomous relative to observations. 
Moreover, scanning the phase space, our stability maps express the variations 
of two explicit parameters, without implicit and external relations.}

\begin{figure*}[!t]
   \centering
     \includegraphics[angle=270,scale=0.37]{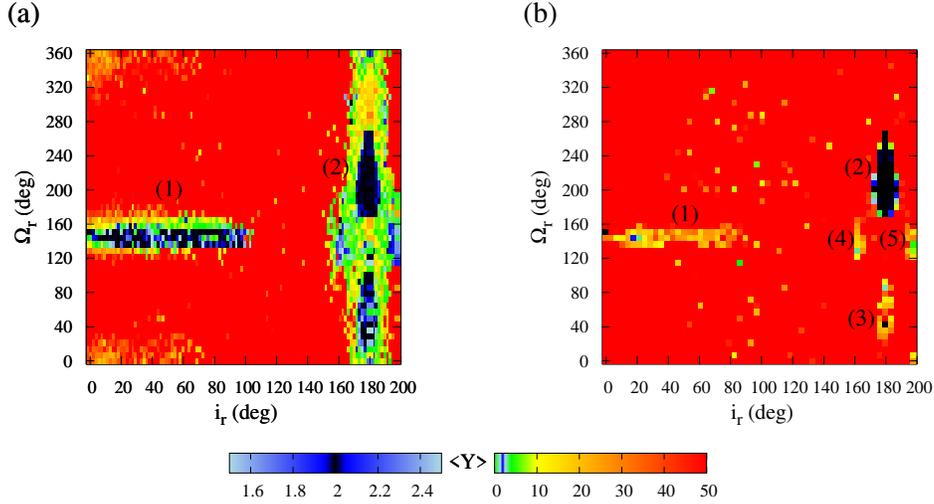}
 \caption{Stability maps for the HD$\thinspace$73526 planetary system 
     in the $[i_r,\Omega_r]$ parameter space. Panel (a) is
     plotted with an integration time of $5000$ yrs while
       planel (b) is plotted over one order of magnitude longer, 
      namely 50000 years. 
     One observes the disappearance of the stable island (1).
     Only the island (2) survives for longest timescales.
     Initial conditions come from the best fit 
     of Tinney \etal (2006) (see Table \ref{tab_param}). 
     Relative inclinations and relative longitudes of nodes are defined as 
     follows~: $i_r=i_c-i_b$ and $\Omega_r=\Omega_c-\Omega_b$ 
     (by convention, at $t=0$, $\Omega_b=0^\circ$ and $i_b=0^\circ$).   
     Resolution of the grid is 110x50 for panel (a), 55x50 for panel (b).
     Black and dark-blue colors represent highly stable orbits 
     ($<$Y$>$ $=2 \pm 3\%$ and $<$Y$>$ $=2 \pm 5\%$ respectively, $<$Y$>$
     being the MEGNO indicator value) while warm colors refer to 
     highly unstable orbits
     ($<$Y$>$ $\gg 2$).}
         \label{fig1} 
\end{figure*}

By applying the observational data of the
HD$\thinspace$73526 planetary system (see Table \ref{tab_param})\footnote{with
in addition at $t=0$, $\Omega_b=\Omega_c=0$, $i_b=0$, and $i_c=1^\circ$
(because of gravitational interactions of the whole 3-body problem, the
relative inclination $i_r=i_c-i_b\neq0$ is then free to evolve in a 3-D
space).}, and  
scanning the non-determined elements, namely the $i_r$ relative inclination 
($i_r=i_c-i_b$) and the $\Omega_r$ relative longitude of nodes 
($\Omega_r=\Omega_c-\Omega_b$), we find two main islands of stability, as shown 
in Fig. \ref{fig1}a. The first (1) is\linebreak obtained for 
$i_r \in [8^\circ,97^\circ]$, and the other (2) for very high relative
inclinations, 
namely $i_r \in [173^\circ,187^\circ]$. In this stability map, we highlight
that stability does not allow coplanar prograde\linebreak orbits. 

\begin{table*}[!ht]
\begin{center}   
   \caption{\label{tab_param}Orbital parameters of the HD$\thinspace$73526
     and HD$\thinspace$160691 planetary systems. Data come from Tinney \etal 
     2006 and McCarthy \etal 2004 respectively.}
   \begin{tabular}{cccccccc}
   \hline
   Planets&
   $\begin{array}{c}
      M_{star}\\
      (M_{\odot})
   \end{array}$&
   $\begin{array}{c}
      m_P \textrm{ sin } i_l\\
      (M_J)
   \end{array}$&
   $\begin{array}{c}
      P \\
      \textrm{(days)}
   \end{array}$&
   $\begin{array}{c}
      a \\
      \textrm{(AU)}
   \end{array}$&
   $\begin{array}{c}
      e \\
      \textrm{} 
   \end{array}$&
   $\begin{array}{c}
      \omega \\
      \textrm{(deg)}
   \end{array}$&
   $\begin{array}{c}
      M \\
      \textrm{(deg)}
   \end{array}$
   \tabularnewline
   \hline

   $\begin{array}{c}
      \textrm{HD$\thinspace$73526 b}\\
      \textrm{HD$\thinspace$73526 c}
   \end{array}$&
   $\begin{array}{c}
   1.08 \pm 0.05
   \end{array}$&
   $\begin{array}{c}
   2.9 \pm 0.2\\
   2.5 \pm 0.3
   \end{array}$&
   $\begin{array}{c}
   188.3 \pm 0.9\\
   377.8 \pm 2.4
   \end{array}$&
   $\begin{array}{c}
   0.66 \pm 0.01\\
   1.05 \pm 0.02
   \end{array}$&
   $\begin{array}{c}
   0.19 \pm 0.05\\
   0.14 \pm 0.09
   \end{array}$&
   $\begin{array}{c}
   203 \pm 9\\
   13 \pm 76
   \end{array}$&
   $\begin{array}{c}
   86 \pm 13\\
   82 \pm 27
   \end{array}$
   \tabularnewline
   \hline

   $\begin{array}{c}
      \textrm{HD$\thinspace$160691 b}\\
      \textrm{HD$\thinspace$160691 c}
   \end{array}$&
   $\begin{array}{c}
   1.08 \pm 0.05
   \end{array}$&
   $\begin{array}{c}
   1.67 \pm 0.11\\
   3.10 \pm 0.71
   \end{array}$&
   $\begin{array}{c}
   645.5 \pm 3\\
   2986 \pm 30
   \end{array}$&
   $\begin{array}{c}
   1.50 \pm 0.02\\
   4.17 \pm 0.07
   \end{array}$&
   $\begin{array}{c}
   0.20 \pm 0.03\\
   0.57 \pm 0.1
   \end{array}$&
   $\begin{array}{c}
   294 \pm 9\\
   161 \pm 8
   \end{array}$&
   $\begin{array}{c}
   0 \\
   12.6 \pm 11.2
   \end{array}$
   \tabularnewline
   \hline

\end{tabular}
\end{center}
\end{table*}

The purpose of fast-chaos indicators, and in particular of MEGNO, is to
predict dynamical behavior over a long timescale using short
integration times. Our integration times do not
mean {\it stability times}  or {\it prediction limits} but, using the MEGNO\linebreak
indicator, they express the minimal times for knowing
  trajectory future. As a
result of the principle of MEGNO, stability is generally
acquired for timescales far longer than the
integration times. The ratio of ``prediction time'' to ``integration time''
achieved by MEGNO, is optimal.

The MIPS maps presented in this paper were confirmed by a second 
global analysis technique (e.g. Marzari \etal  2006), 
based on Laskar's (1993) Frequency Map Analysis (FMA).
The FMA method uses the diffusion rates
of intrinsic frequencies as a measure for stability. The numerical values
of these frequencies are provided by this method.
The lowest intrinsic frequencies determine the necessary integration time.
It is, therefore, possible that the FMA method requires longer integration
times than MEGNO. 

While the necessary integration time for FMA is provided by the lowest
intrinsic frequency, one is a priori free to choose the time when applying 
MEGNO.
The advantage is that a shorter integration time can be used. On the other 
hand, there is an\linebreak uncertainty about the good choice of the integration time
which might be too short. We, therefore, produce maps at different\linebreak instant of
times and consider the evolution of the most stable regions.
Fig. \ref{fig1}a and Fig. \ref{fig1}b show newly-obtained $[i_r,\Omega_r]$ 
maps for the HD$\thinspace$73526 system at 5000 and 50000 years respectively.
While the first island (1) of Fig. \ref{fig1}a completely disappears
in\linebreak Fig. \ref{fig1}b, the second one remains always highly stable. Continuing
in time, island (2) persists.
We also find three very small stable islands (3), (4) and (5). One of the
islands (3) is separated by $180^\circ$ in 
$\Omega_r$ from the center of the large island (2). The two other islands, 
(4) and (5), are distributed symmetrically with respect to the islands
(2) and (3). As a consequence, due to the lifetime and size of each
stability zone, the large island (2) with
$i_r \in [173^\circ,187^\circ]$ and $\Omega_r \in [173^\circ,266^\circ]$
contains the most stable orbits (i.e. the least ``model dependent'' on added
perturbations).\footnote{Let us note that a
relative inclination around $180^\circ$ is equivalent to a planar problem
where one planet has a retrograde motion  
with respect to the other. 
Therefore, considering a scale change of $180^\circ$ in relative 
inclinations, we will use the notation $i_r^{retro}=1^\circ$ instead of 
$i_r=179^\circ$, in the following.}
This does not necessarily imply that the two observational\linebreak
planets of the HD\thinspace73526 system are {\it counter-revolving
  planets}.\footnote{Counter-revolving planets mean that the orbital elements 
of the two planets are orbiting in opposite directions about the central
star.} As a consequence, we study in the following section
whether such a counter-revolving configuration is consistent with
observational data.

\section{Observational data fits}\label{obs}
Error-bars in published, orbital elements are significant and a fit 
including new observations may yield quite different orbital\linebreak
elements. By using different orbital elements that are consistent with 
observational data, S\'andor \etal (2007) found stability for
the coplanar and prograde case. On the one hand, for lowest\linebreak values of 
$\chi^2$ and $rms$ ( $\chi^2=1.57$ and $rms=7.9$ $m.s^{-1}$ for the best 
dynamical fit of Tinney \etal 2006), we find a weak chaotic solution. On the 
other hand, S\'andor \etal (2007) obtain four\linebreak stable solutions but with values
of $\chi^2$ and $rms$ somewhat higher than those of Tinney ($\chi^2 \in
[1.58;1.87]$ and $rms \in [8.04;8.36]$). As a consequence, the right
astrometric characterization of the HD$\thinspace$73526 planetary system still remains open.

\begin{figure}[!h]
   \centering
     \includegraphics[scale=0.45]{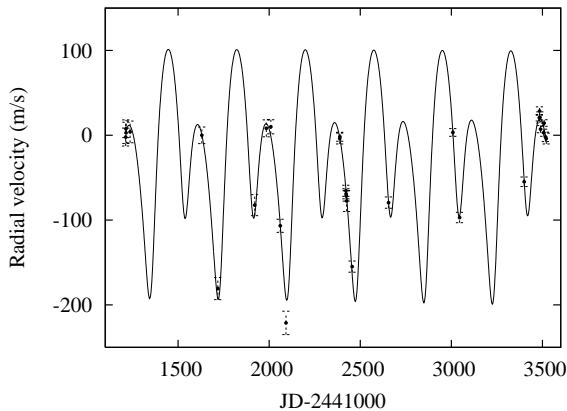}
   \caption{Dynamical velocity fit with measured velocities of the
     HD\thinspace73526 planetary system. The best
     dynamical fit leading to a stable two-planet system is obtained for : 
     $m_b=2.4921$ $M_{Jup}$, $m_c=2.5919$ $M_{Jup}$,
     $P_b=187.935$ days, $P_c=379.795$ days, $a_b=0.6593$ AU, $a_c=1.0538$ AU,
     $e_b=0.2401$, $e_c=0.2048$, $i_b=0^\circ$, $i_c=180^\circ$, 
     $\Omega_b=0^\circ$, $\Omega_c=0^\circ$, $\omega_b=184.569^\circ$, 
     $\omega_c=58.545^\circ$, $M_b=97.297^\circ$, and $M_c=221.361^\circ$. 
     The velocity offset is : $V_0=-25.201\, m.s^{-1}$. The corresponding
     $rms$ residuals 
     are $6.34\, m.s^{-1}$ while the $\chi^2$ reduced factor
     is equal to $1.257$.}
   \label{fig2}   
\end{figure}

We have performed orbital fits for counter-revolving configurations using a
genetic algorithm (called Pikaia; see Charbonneau 1995) based on a fitting
method.\footnote{Orbital fitting process are notably explained 
in Beaug\'e \etal (2007).} We find stable 
retrograde solutions for values of $\chi^2$ and $rms$ smaller than the
prograde fits of Tinney \etal (2006) and S\'andor \etal (2007) 
($\chi^2=1.257$ and $rms=6.34$ $m.s^{-1}$). The radial velocity curve of the
best stable fit is shown in 
Fig. \ref{fig2}. It is very similar to the radial velocity
curve given by Tinney \etal (2006) and S\'andor \etal (2007).
As a consequence, we point out that the possibility of counter-revolving
planets should not be discarded in observational-data fits.

Nevertheless, whatever the directions of motions of the two planets are, the 
$\chi^2$ values are significantly above the expected value of $1.0$. More 
observations would enable better fits to be derived. However in these
conditions, the possibility may not be excluded that the HD\thinspace73526
planetary system is a counter-revolving system. 
Anyway, the counter-revolving configuration related to the HD\thinspace73526
planetary system is consistent with the observational data. 
From a dynamical point of view, counter-revolving 
orbits are all the more plausible because they have larger highly stable
regions.

\section{Statistical approach}\label{stat}
By integrating 1000 random systems (according to a Monte-Carlo method) within 
the error-bars proposed by Tinney \etal (2006), we obtain statistically more
stable solutions for coplanar counter-revolving orbits than for prograde 
ones. For prograde coplanar orbits, we find only 17 stable systems while for 
counter-revolving coplanar orbits, we obtain 500 stable systems. 

The occurence of stable counter-revolving systems also\linebreak appears in the 
neighborhood of other two-planet systems. The statistical results for their 
stability in the prograde case and in the counter-revolving one are 
presented in Table \ref{tab_stat} for two\linebreak additional 2:1 and two 5:1 resonance
cases. In all cases, a significant number of stable systems in retrograde
resonances is found. The high statistical occurence of stable
retrograde configurations justifies the study of such solutions, whether or not
they correspond at present to observational data.
On the other hand, we are well aware of the cosmogonic problem
for obtaining planets\linebreak in retrograde resonances within the frame
of current theories of planetary formation. During the early dynamical 
evolution of planetary systems, planets may end up on retrograde orbits (as
discussed in the following section). 

\begin{table}[!ht]
\begin{center}
   \begin{tabular}{cccc}
   \hline
   System sources&
   Period ratio&
   $\begin{array}{c}
   \textrm{Prograde}\\
   \textrm{MMR}
   \end{array}$&
   $\begin{array}{c}
   \textrm{Retrograde}\\
   \textrm{MMR}
   \end{array}$
   \tabularnewline
   \hline
  
   HD73526&
   2/1&
   17&
   500\tabularnewline

   HD82943&
   2/1&
   755&
   1000\tabularnewline

   HD128311&
   2/1&
   249&
   137\tabularnewline

   HD160691&
   5/1&
   0&
   320\tabularnewline

   HD202206&
   5/1&
   0&
   631\tabularnewline
   \hline
\end{tabular}
   \caption{\label{tab_stat} Statistical results about possibility of stable
     systems to be in retrograde resonance. 1000 random systems have been 
     integrated in their errors bars and assuming prograde coplanar orbits or
     retrograde ones. The number of stable systems is indicated in each case.
     Data come from Tinney \etal (2006), Mayor \etal (2004),
     Vogt \etal (2005), McCarthy \etal (2004) and Correia \etal (2005; table 4)
     respectively.} 
\end{center}
\end{table}

\section{Formation of counter-revolving planets}\label{form}
Up to now, few works have been carried out on the
formation\linebreak of highly-inclined or counter-revolving orbits. For instance, 
Thommes \& Lissauer (2003) showed that a planetary migration leading to 
resonance capture in 2:1 MMR may cause a significant increase in the mutual 
inclinations of the planets. However, the resulting configuration
never seems to exhibit retrograde\linebreak motions. That is why, in 
this section, we propose two novel mechanisms of formation of
counter-revolving configurations.

It has been known for a few years that free-floating planetary-mass objects
have been located in interstellar space (see for instance Zapatero Osorio \etal
2000 and Lucas \etal 2005). These
free-floating planets may interact with planetary systems in their host
cluster; they are either 
scattered or captured. The recent studies of Varvoglis (2008) show that by 
integrating the trajectories of planet-sized bodies that 
encounter a coplanar, two-body system (a Sun-like star and a Jupiter mass), 
the probability of capture is significant, and almost half of 
the temporary captures are found to be of the counter-revolving type. 
Although captures of 
free-floating planets remain speculative, this could be a 
feasible mechanism for generating counter-revolving orbits.

\begin{figure*}[!th]
   \centering
     \includegraphics[angle=270,scale=0.38]{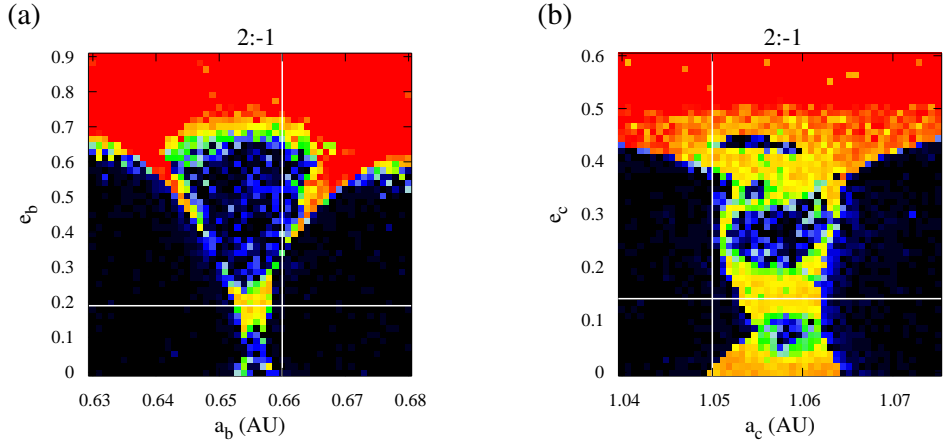}
   \caption{Stability maps in the $[a_b , e_b]$ and $[a_c, e_c]$
     parameter spaces based on the HD$\thinspace$73526 planetary system, 
     taking into account initial conditions $(C)$ ($i_r^{retro}=1^\circ$ and 
     $\Omega_r=216^\circ$). Color scale is 
     the same as in Fig. \ref{fig1}. Resolution of the grid is 50x50. The 
     pseudo-observational point is indicated by the intersection of horizontal 
     and vertical lines. V-shape structures correspond to the 2:1 retrograde
     MMR (2:-1 MMR).} 
   \label{fig3}   
\end{figure*}

Forming close-in planets by using the slingshot model\linebreak revisited by
Nagasawa \etal (2008) is another possibility. 
Starting from a hierarchical 3-planet system and considering\linebreak a 
migration mechanism including process of planet-planet scattering and tidal 
circularization, the authors show indeed that close-in
planets may be formed. In a few cases, due to the Kozai mechanism (involving
exchanges between eccentricities and\linebreak inclinations), one planet enters a
retrograde motion. 

Considering (1) these two feasible mechanisms of formation\linebreak of
counter-revolving orbits, (2) the dynamical fit obtained in Section
\ref{obs}, and (3) the statistical occurence of retrograde solutions
(Section \ref{stat}), we may say that counter-revolving
theoretical\linebreak configurations are serious candidates 
for real systems (that could be observed later). 
In the future, we will study these two formation
processes more deeply. In the following sections, we focus\linebreak on the
specific dynamical
behavior of systems harboring counter-revolving planets.

\section{Fine structure of the resonance}\label{structure}
Due to the retrograde motion of planet c\footnote{By
convention, the orbital plane of the planet b is chosen as a
reference plane. The planet b is supposed moving in the prograde direction 
while the planet c in the retrograde direction. Results remain analoguous 
with the reverse assumption (b retrograde and c prograde).}, 
the orbital resonance\linebreak of a given planetary system (e.g. the 
HD$\thinspace$73526 planetary\linebreak system) is called a 2:1 retrograde MMR 
(that we also annotate as a 2:-1 MMR). Studying the fine structure of this MMR
provides\linebreak 
a key to understand the stability of the system. We assume\linebreak 
initial conditions taken from the stability island (2) of
the $[i_r,\Omega_r]$ map (Fig. \ref{fig1}b), that is to say 
the elements of Table \ref{tab_param} and in\linebreak addition \,
$\Omega_r~=~216^\circ$ and
$i_r^{retro}=1^\circ$. This set of initial conditions is noted (C).\footnote{For $\Omega_r=216^\circ$ and
$i_r^{retro}=1^\circ$, the value of $\chi^2$ obtained with the velocity offset
$V_0=-38$  $m.s^{-1}$ is $2.44$. The rms residuals are $12.31$.}. We then
obtain the two $[a_b,e_b]$ and $[a_c,e_c]$\linebreak stability maps presented in
Fig. \ref{fig3}. The presence of a strong MMR generates clear instability
zones with a prominent V-shape structure in Fig. \ref{fig3}a. We note the
narrowness of the V-shapes,\linebreak approximately $0.006$ AU wide for the inner orbit
(when $e_b=0$) and $0.0015$ AU wide for the outer one (when $e_c=0$). As
a comparison, V-shape structures of the Sun-Jupiter-Saturn system\linebreak are five times
and twice as large respectively (for Jupiter and Saturn).  
We note in 
addition how the pseudo-observational point lies at the edge of
the V-shapes (Fig. \ref{fig3}).

\section{Evidence for a retrograde resonance}\label{evidence}
Maps for extrasolar-planet systems, with stability regulated by a
prograde MMR, are characterized by small islands of 
stability\linebreak (or linear stable strips in $[a_b,a_c]$ maps) inside large zones of\linebreak
instability. By contrast, in the case of retrograde resonance and when
assuming the initial conditions (C), we detect a dense,\linebreak stable regime in a
series of maps (e.g.  $[a_b,a_c]$), except for one unstable
  zone related to the MMR.
In several cases of resonant prograde systems, it has  been shown that planets 
on highly stable\linebreak orbits may avoid close approaches due to their adequate
positions over their orbits and apsidal line locking (see Bois
\etal 2003). This mechanism of stability is not lost during the
dynamical evolution of the system when the apsidal lines on average precess at
the same rate (i.e. the ASP phenomenon). 
Without such a protection mechanism of 2:1 MMR combined to 
an ASP, disturbing close approaches between the planets are
theoretically possible in various planet positions. In the case of the 2:1
retrograde MMR, one  
planet being retrograde, orbital motions occur then in opposite 
directions. As a consequence, the length of time that planets spend in
conjunction, is much shorter for counter-revolving orbits than for prograde
ones. 
This could explain the narrowness of the V-shapes in Fig. \ref{fig3}.

In Gayon \& Bois (2008), we showed that, in cases of very compact planetary
systems obtained by a scale reduction of a given observed system, 
``retrograde'' stable islands survive, in contrast to ``prograde'' ones that
disappear.  
This scale reduction and the V-shapes of Fig. \ref{fig3} illustrate the
efficiency of retrograde MMRs for providing stability.
In addition, this MMR mechanism is coupled to specific behaviors of the 
apsidal lines, as shown in the following section.

\section{A new mechanism of apsidal precession at retrograde resonance}\label{apsidal}
In the case of the 2:1 retrograde MMR (2:-1 MMR), the expressions for the 
resonance angles $\theta_1$ and $\theta_2$, and the ASP angle 
$\theta_3$ are~:
\begin{center}
\begin{tabular}{l}
$\theta_1 = - \lambda_b - 2 \lambda_c + 3 \tilde{\omega}_b$\\
$\theta_2 = - \lambda_b - 2 \lambda_c + 3 \tilde{\omega}_c$\\
$\theta_3 = 3 \,(\tilde{\omega}_b - \tilde{\omega}_c)$
\end{tabular}
\end{center}
\newpage
\noindent
where $\lambda$ is the mean longitude, and $\tilde{\omega}$ is the apsidal
longitude, defined by $\lambda=M+\tilde{\omega}$\, and \,$\tilde{\omega} =
\Omega + \omega$ respectively, 
for prograde orbital motion of a planet; for retrograde
motion, these variables are defined to be:
$\lambda=-M+\tilde{\omega}$ and $\tilde{\omega}  = \Omega - \omega$.
The
general expression for the relative apsidal longitude is:
$\Delta{\tilde{\omega}}=\tilde{\omega}_b - \tilde{\omega}_c=\theta_3/q$,
where $q$ is the order of the resonance. 
A more thorough investigation of retrograde
resonances, using an analytical approach, is in preparation (Gayon \etal 2008).

 \begin{figure}[!h]
\vspace{2mm}
     \centering
     \includegraphics[scale=0.38]{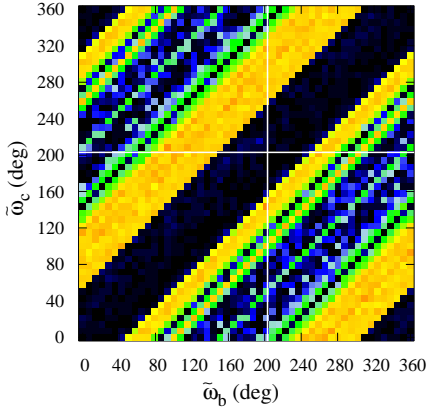}
     \caption{ Stability map in the $[\tilde{\omega}_b,\tilde{\omega}_c]$ 
     parameter space based on the HD$\thinspace$73526 planetary system taking 
     into account initial conditions $(C)$
     ($i_r^{retro}=1^\circ$ and $\Omega_r=~216^\circ$). The
     pseudo-observational point is inside the stability linear strip.
     Color scale is the same as in Fig. \ref{fig1}. Resolution of the grid is 
     50x50.}
   \label{fig4} 
\end{figure}

\begin{figure}[!ht]
   \centering
     \includegraphics[scale=0.28]{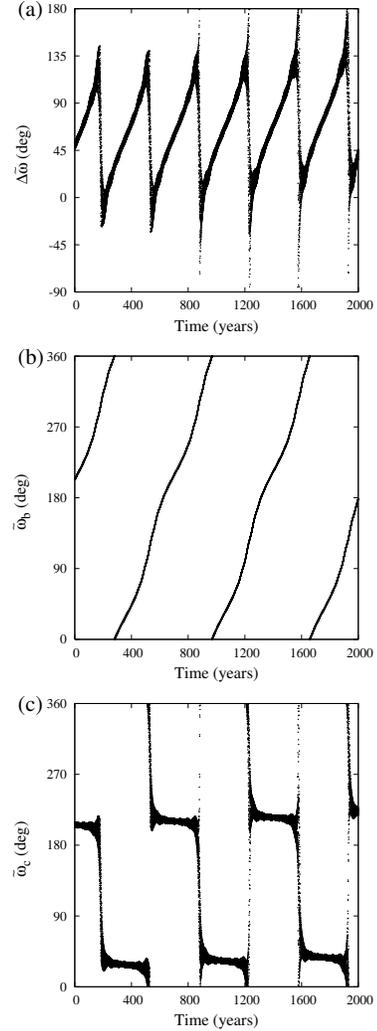}
   \caption{Time variation of the $\Delta{\tilde{\omega}}$ angle (a),
     $\tilde{\omega}_b$ (b), and $\tilde{\omega}_c$ (c)  for initial 
     conditions $(C)$. In panel (b), the slope of 
     $\tilde{\omega}_b$ is positive while the one of $\tilde{\omega}_c$ 
     (panel (c)) is negative. Panel (a) expresses the behavior
     of the $\Delta\tilde{\omega}$ combination of $\tilde{\omega}_b$ and 
     $\tilde{\omega}_c$ (see text).}
   \label{fig5}   
\end{figure}

In Fig. \ref{fig4} we plot the $[\tilde{\omega}_b , 
\tilde{\omega}_c]$ parameter space that shows a stable linear strip,
in dark-blue, including 
the ``pseudo-observational'' point. We learn
that stable solutions are possible\linebreak only when $\tilde{\omega}_b$ and 
$\tilde{\omega}_c$ precess, on average, at the same rate. The
stabilizing mechanism of the system involves a synchronous\linebreak precession of the
apsidal lines. The two longitudes of periastron 
do not precess however in the same direction. The outer orbit is affected by a 
retrograde precession $(-\tilde{\omega}_c)$ relative to the inner\linebreak orbit's 
precession $(\tilde{\omega}_b)$ (see Fig. \ref{fig5}c {\it vs} \ref{fig5}b).
Writing the longitudes\linebreak of periastron as directed angles, we
find that the relative apsidal\linebreak longitude $\Delta\tilde{\omega}$ neither
circulates nor librates clearly, as shown in Fig. \ref{fig5}a.
$\Delta{\tilde{\omega}}$ presents a strange motion composed of~: (1) a phase 
of prograde circulation including librations with amplitudes of $\pm
8^\circ$, then following a sharp reversal of circulation direction, (2)
a phase of fast retrograde circulation, until a second, sharp\linebreak reversal. These 
two phases alternate successively according to alternations (or rocking) of
$180^\circ$, which correspond to a sort of 
cusp.\footnote{At phase transitions, a scattering of dots
appears. When the outer-orbit eccentricity goes to zero,  
the $\tilde{\omega}_c$ angle that depends on the\linebreak $(a-r)/ae$ ratio is not 
defined.} We note that $\tilde{\omega}_c$, in contrast to the case for 
$\tilde{\omega}_b$, does not {\it uniformly} circulate but presents
retrograde circulation phases\linebreak interrupted with short libration intervals (
Fig. \ref{fig5}c {\it vs} \ref{fig5}b).
In spite of the opposite directions of their precession, both orbits\linebreak precess,
on average, at the same rate. As a consequence, the planetary
system is affected by an apsidal synchronous precession.\linebreak
Considering the alternating behavior of the $\Delta\tilde{\omega}$ 
angle or the\linebreak unusual presence of cusp in
the $\Delta{\tilde{\omega}}$ behavior, we refers to this new 
stabilizing factor as either an {\it alternating} ASP, or 
a {\it rocking}\linebreak ASP (or RASP). We have produced movies
illustrating the mechanisms introduced in the present paper, in
particular the phenomenon of RASP.\footnote{Movies are
downloadable from:\\ http://www.oca.eu/gayon/Extrasolar/Retro\_MMR/movies.html}

\begin{figure}[!t]
   \centering
   \includegraphics[scale=0.38]{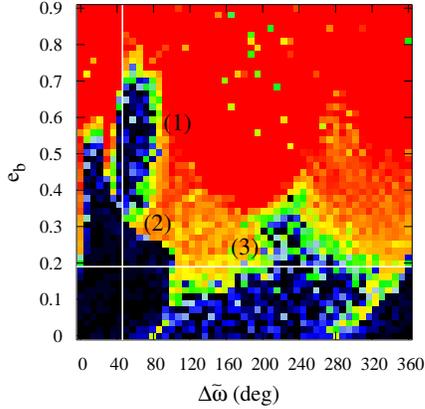}
   \caption{Stability map in the $[\Delta{\tilde{\omega}},e_b]$ parameter 
     space based on the HD$\thinspace$73526 planetary system, taking into 
     account initial conditions $(C)$ (Table \ref{tab_param} with in addition 
     $i_r^{retro}=1^\circ$).
     The pseudo-observational point is inside the stability large island (2) 
     characterized by a 2:1 retrograde near-resonance and an alternating ASP.
     Color scale is the same as in Fig. \ref{fig1}. Resolution of the grid is
     50x50.}
   \label{fig6}   
\end{figure}

Relations between the eccentricity of the inner orbit and the 
$\Delta{\tilde{\omega}}$ angle exist. In $[\Delta{\tilde{\omega}},e_b]$
parameter space, we can see from Fig. \ref{fig6} how the
2:1 retrograde MMR spreads out its resources:
 
(1) Inside the 2:-1 MMR (i.e. inside the $[a,e]$ V-shape), both apsidal lines 
on average precess at the same rate and in the same prograde direction. The 
2:1 retrograde MMR is then combined with a uniformly {\it prograde} ASP
(island (1)).  

(2) Close to the 2:-1 MMR (i.e. outside but close to the $[a,e]$ V-shape), both 
apsidal lines on average precess at the same rate but in opposite directions. 
The 2:1 retrograde {\it near}-MMR is combined with the mechanism of {\it
  rocking} ASP (island (2)).  

(3) The division between these two islands is related to the degree 
of closeness to the 2:1 retrograde MMR. We highlight the fact that, for a long 
timescale, the third island in the  
$[\Delta{\tilde{\omega}},e_b]$ map, is proved to be a {\it chaotic} zone 
(island (3) where $\Delta{\tilde{\omega}} \in [80^\circ,280^\circ]$).  

We note that a mechanism of 
stability involving an ASP may persist far from the MMR in the 
prograde case, while it disappears for short distance to the MMR in the
counter-revolving case. Hence, moving away from the $[a, e]$ V-shape of the
2:-1 MMR (Fig. \ref{fig3}), we find that both apsidal lines precess in
opposite\linebreak directions but at different rates.

 By studying the parameter-space in the vicinity of the best-fit of the 
HD$\thinspace$73526 planetary-system and searching for stable\linebreak configurations 
with similar values of {\it rms}, a new, theoretical mechanism of stability has 
been discovered. It is characterized by a 2:1 retrograde, near-MMR combined to 
a rocking ASP. 
Such a stability also allows a large range of eccentricities
(see for instance planet b in Fig. \ref{fig6}). Such a mechanism 
is particularly robust on large
timescales. It is why such mechanisms involving such resources of the
2:-1 orbital resonance could prove to be\linebreak
relatively generic and suitable for the 
stability of a class of\linebreak compact multi-planetary systems where other solutions
of the 3-body problem are not possible.

\section{The 5:1 retrograde MMR}\label{HD16}
The mechanism of stability involving both a retrograde MMR and an ASP is also
found for the 5:1 orbital period ratio by scanning initial
conditions in the vicinity of
the HD$\thinspace$160691 planetary system (McCarthy \etal 2004, planets b and
c, see Table \ref{tab_param}). By detecting a fourth planet, we note that new 
observations of this system have modified the orbital structure of the
entire system (Pepe \etal 2007). The observations of Pepe
\etal (2007) show furthermore that a new coplanar fit of prograde orbits and a new fourth
planet seems to solve the problem without the need for retrograde
resonance. We note that in the vicinity of the best fit
solution of McCarthy \etal (2004), it is possible to find an\linebreak 
example of 5:1 retrograde resonance. We consider this possibility 
as an academic investigation of this order
  of MMR.

\begin{figure}[!t]
   \centering
   \includegraphics[scale=0.38]{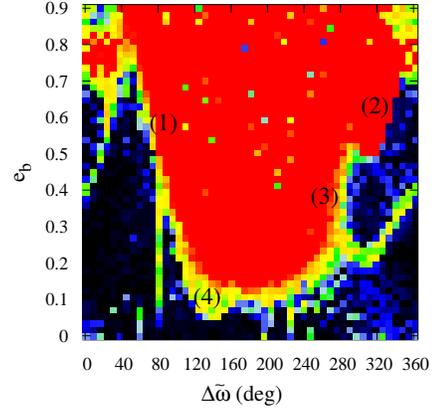}
   \caption{Stability map in the $[\Delta{\tilde{\omega}},e_b]$ parameter 
     space based on the HD$\thinspace$160691 planetary system (data from 
     McCarthy \etal (2004) shifted to the 5:1 retrograde MMR (i.e. $a_b=1.44$ 
     AU), see Table \ref{tab_param} with in addition $i_r^{retro}=1^\circ$). 
     Islands (1) and (3) are characterized by a 5:-1 
     MMR and an ASP with apsidal alignement while islands (2) and (4) by an 
     5:-1 MMR and a circulation of the $\Delta\tilde{\omega}$ variable. Color 
     scale is the same as in Fig. \ref{fig1}. Resolution of the grid is 50x50.}
   \label{fig7}   
\end{figure}

The important point is that this 3-body system is completely unstable
for prograde orbits. Nevertheless, by scanning the non-determined parameter
space, only one island of stability exists (for $i_r \in
[156^\circ,204^\circ]$ and $\Omega_r \in [7^\circ,180^\circ]$) and after
an analysis we highlight the retrograde motion of the outer planet. Moreover,
scanning the $[\Delta\tilde{\omega},e_b]$ parameter space of the system in
5:1 retrograde MMR, as for the HD$\thinspace$73526 planetary system in 2:1
retrograde MMR, we observe the distribution of the different 
apsidal behaviors (see Fig. \ref{fig7}). Islands (1) and (3) are characterized
by an ASP with 
an  apsidal alignement. More precisely, both longitudes of periastron
($\tilde{\omega}_b$ and $\tilde{\omega}_c$) on average precess at the same 
rate, both in the retrograde direction, while the $\Delta\tilde{\omega}$
variable\linebreak 
librates about $0^\circ$: it is a uniformly {\it retrograde} ASP.  Within\linebreak
islands (2) and (4), $\tilde{\omega}_b$ and $\tilde{\omega}_c$ precess in
opposite directions but according to different rates: $\Delta\tilde{\omega}$
circulates. Hence, no apsidal line locking is required for stability. 
We find fine, V-shape structures in $[a,e]$ maps corresponding 
to the 5:1 retrograde MMR. The width is 0.02 AU for the inner planet (when
$e_b=0$) and only 0.002 AU for the outer one (when $e_c=0$). Consequently, the
dynamical study of this case derived from the HD$\thinspace$160691 planetary
system allows us to find another theoretical possibility of
stability involving resources of a retrograde MMR. 

\section{Conclusion}
We have found novel mechanisms giving rise to stability that could be
suitable for a class of compact planetary systems. Such mechanisms involve
counter-revolving orbits forming a retrograde MMR occuring in a
quasi-identical plane. High statistical\linebreak occurence of stable
counter-revolving orbits is found.
Our study of retrograde MMRs indicates the large stability domains and
the specific behaviors of the precession and resonant angles.
We propose that these large stability domains are caused by 
close approaches much faster 
and shorter for counter-revolving\linebreak configurations than for the prograde ones. 
Scanning the 
HD\thinspace73526 planetary system, we find evidence for a new type of apsidal
precession (the rocking ASP). 
We find that the\linebreak difference between the longitudes of 
periastron reveals a specific alternation mode at retrograde resonances.  
We emphasize that the
counter-revolving configuration studied for the HD\thinspace73526 planetary
system is consistent with the observational data.
Free-floating planets or the Slingshot model might explain the origin
of such counter-revolving systems.

\begin{acknowledgements}
We thank Alessandro Morbidelli, Makiko Nagasawa, Hans Scholl, 
and Charalampos Varvoglis for useful discussions.
We particularly thank Cristi\'an Beaug\'e for providing
us a code performing orbital  
fits and, as a referee, for very constructive comments that greatly helped to
improve the paper.  
\end{acknowledgements}

\end{document}